\documentclass[trackchanges, twocolumn]{aastex701}
\usepackage{float}
\usepackage{graphicx}
\usepackage{hyperref}
\usepackage{xspace}
\usepackage{xcolor}
\newcommand\kms{\ensuremath{{\rm km\,s}^{-1}}\xspace}
\newcommand\satname{DELVE~8/Gemini~I\xspace}

\newcommand{\CHECK}[1]{#1}

\newcommand{\update}[1]{{#1}}

\newcommand{\satmv}{$M_V = -2.1^{+0.4}_{-0.6}$}
\newcommand{\satmvnolabel}{$-2.1^{+0.4}_{-0.6}$}
\newcommand{\satmstar}{$M_* = 1180^{+820}_{-320} \ \rm M_\odot$}
\newcommand{\satlv}{$L_V = 590^{+410}_{-160}\ \rm L_\odot$}
\newcommand{\satmod}{$(m-M)_0 = 20.39^{+0.07}_{-0.06} \pm 0.1$ (sys.)}
\newcommand{\satheliodist}{$D_\odot = 120_{-6}^{+7}~{\rm kpc}$}

\defcitealias{McQuinn:2025hst..prop18066M}{HST GO Cycle 33 \#18066}

\reportnum{FERMILAB-PUB-25-0714-LDRD-PPD}

\begin{document}

\title{ \update{Discovery and  Spectroscopic Characterization of a  Distant, Compact Milky Way Satellite in Gemini}}

\correspondingauthor{\\Katherine Overdeck (\url{koverdeck@uchicago.edu}), \\
William Cerny (\url{william.cerny@yale.edu}), \\ Chin Yi Tan  (\url{chinyi@uchicago.edu})
}

\author[0009-0008-0959-0162]{K.~Overdeck}
 \email{koverdeck@uchicago.edu}
\affiliation{Kavli Institute for Cosmological Physics, University of Chicago, Chicago, IL 60637, USA}
\affiliation{Department of Astronomy and Astrophysics, University of Chicago, Chicago, IL 60637, USA}
\affiliation{NSF-Simons AI Institute for the Sky (SkAI),172 E. Chestnut St., Chicago, IL 60611, USA}

\author[0000-0003-1697-7062]{W.~Cerny}
\email{william.cerny@yale.edu}
\affiliation{Department of Astronomy, Yale University, New Haven, CT 06520, USA}

\author[0000-0003-0478-0473]{C.~Y.~Tan}
\email{chinyi@uchicago.edu}
\affiliation{Kavli Institute for Cosmological Physics, University of Chicago, Chicago, IL 60637, USA}
\affiliation{Department of Physics, University of Chicago, Chicago, IL 60637, USA}
\affiliation{NSF-Simons AI Institute for the Sky (SkAI),172 E. Chestnut St., Chicago, IL 60611, USA}

 \author[0000-0002-9144-7726]{C.~E.~Mart\'inez-V\'azquez}
  \email{clara.martinez@noirlab.edu}
 \affiliation{NSF NOIRLab, 670 N. A'ohoku Place, Hilo, Hawai'i, 96720, USA}

\author[0000-0002-6021-8760]{A.~B.~Pace}
\email{pvpace1@gmail.com}
\affiliation{Department of Astronomy, University of Virginia, 530 McCormick Road, Charlottesville, VA 22904, USA}

\author[0009-0001-1133-5047]{J.~A.~Sharp}
\affiliation{Kavli Institute for Cosmological Physics, University of Chicago, Chicago, IL 60637, USA}
\affiliation{Department of Astronomy and Astrophysics, University of Chicago, Chicago, IL 60637, USA}
  \email{sharpj@uchicago.edu}

\author[0000-0001-8251-933X]{A.~Drlica-Wagner}
\email{kadrlica@fnal.gov}
\affiliation{Fermi National Accelerator Laboratory, P.O.\ Box 500, Batavia, IL 60510, USA}
\affiliation{Kavli Institute for Cosmological Physics, University of Chicago, Chicago, IL 60637, USA}
\affiliation{Department of Astronomy and Astrophysics, University of Chicago, Chicago, IL 60637, USA}
\affiliation{NSF-Simons AI Institute for the Sky (SkAI),172 E. Chestnut St., Chicago, IL 60611, USA}

\author[0000-0002-7007-9725]{M.~Geha}
\email{marla.geha@yale.edu}
\affiliation{Department of Astronomy, Yale University, New Haven, CT 06520, USA}

\author[0000-0003-4102-380X]{D.~J.~Sand}
\email{dsand@arizona.edu}
\affiliation{Department of Astronomy/Steward Observatory, 933 North Cherry Avenue, Room N204, Tucson, AZ 85721-0065, USA}

\author[0000-0002-3690-105X]{J.~A.~Carballo-Bello}
 \affiliation{Instituto de Alta Investigaci\'on, Universidad de Tarapac\'a, Casilla 7D, Arica, Chile}
 \email{jcarballo@academicos.uta.cl}

 \author[0000-0003-1680-1884]{Y.~Choi}
 \affiliation{NSF NOIRLab, 950 N. Cherry Ave., Tucson, AZ 85719, USA}
 \email{yumi.choi@noirlab.edu}

 \author[0000-0002-1763-4128]{D.~Crnojevi\'c}
\affiliation{Department of Physics \& Astronomy, University of Tampa, 401 West Kennedy Boulevard, Tampa, FL 33606, USA}
\email{dcrnojevic@ut.edu}

\author[0000-0001-6957-1627]{P.~S.~Ferguson}
 \email{pferguso@uw.edu}
\affiliation{DiRAC Institute, Department of Astronomy, University of Washington, 3910 15th Ave NE, Seattle, WA, 98195, USA}

  \author[0009-0003-2043-5386]{M.~Hirschauer}
 \affiliation{Illinois Math and Science Academy, 1500 Sullivan Ln, Aurora Illinois}
\email{mhirschauer@imsa.edu}

  \author[0000-0001-5160-4486]{D.~J.~James}
 \affiliation{ASTRAVEO LLC, PO Box 1668, Gloucester, MA 01931}
 \affiliation{Applied Materials Inc., 35 Dory Road, Gloucester, MA 01930}
\email{djames44@gmail.com}

\author[0000-0002-3204-1742]{N.~Kallivayalil}
\affiliation{Department of Astronomy, University of Virginia, 530 McCormick Road, Charlottesville, VA 22904, USA}
\email{njk3r@virginia.edu}

 \author[0000-0002-9269-8287]{G.~Limberg}
 \affiliation{Kavli Institute for Cosmological Physics, University of Chicago, Chicago, IL 60637, USA}
 \email{limberg@uchicago.edu}

\author[0000-0002-1763-4128]{P.~Massana}
\affiliation{NSF NOIRLab, Casilla 603, La Serena, Chile}
\email{pol.massana@noirlab.edu}

\author[0000-0003-3519-4004]{S.~Mau	}
\affiliation{Department of Physics, Duke University Durham, NC 27708, USA}
\email{sidney.mau@duke.edu}

\author[0000-0003-0105-9576]{G.~E.~Medina}
\affiliation{Department of Astronomy and Astrophysics, University of Toronto, 50 St. George Street, Toronto ON, M5S 3H4, Canada}
\email{gustavo.medina@utoronto.ca}

 \author[0000-0001-9649-4815]{B.~Mutlu-Pakdil}
 \affiliation{Department of Physics and Astronomy, Dartmouth College, Hanover, NH 03755, USA}
\email{Burcin.Mutlu-Pakdil@dartmouth.edu}

\author[0000-0001-9438-5228]{M.~Navabi}
\email{m.navabi@surrey.ac.uk}
\affiliation{Department of Physics, University of Surrey, Guildford GU2 7XH, UK}

 \author[0000-0002-1793-3689]{D.~L.~Nidever}
 \affiliation{Department of Physics, Montana State University, P.O. Box 173840, Bozeman, MT 59717, USA}
 \email{david.nidever@montana.edu}	

  \author[0000-0002-8282-469X]{N.~E.~D.~No\"el}
 \affiliation{Department of Physics, University of Surrey, Guildford GU2 7XH, UK}
 \email{n.noel@surrey.ac.uk}

 \author[0009-0008-9641-6065]{A.~Pai}
 \affiliation{Kavli Institute for Cosmological Physics, University of Chicago, Chicago, IL 60637, USA}
 \affiliation{Department of Physics, University of Chicago, Chicago, IL 60637, USA}
 \affiliation{NSF-Simons AI Institute for the Sky (SkAI),172 E. Chestnut St., Chicago, IL 60611, USA}

 \email{pai@uchicago.edu}

  \author[0000-0001-5805-5766]{A.~H.~Riley}
 \affiliation{ Lund Observatory, Division of Astrophysics, Department of Physics, Lund University, SE-221 00 Lund, Sweden}
\email{alexander.riley@fysik.lu.se}

\author[0000-0002-1594-1466]{J.~D.~Sakowska}
\email{jsakowska@iaa.csic.es}
\affiliation{Instituto de Astrof\'isica de Andaluc\'ia (CSIC), Glorieta de la Astronom\'ia,  E-18080 Granada, Spain}

\author[0000-0003-3402-6164]{L.~Santana-Silva}
 \affiliation{Observatório do Valongo/UFRJ, Ladeira do Pedro Antônio, 43 - Centro, Rio de Janeiro - RJ, 20080-090, Brazil}
 \email{luidhy@ov.ufrj.br}

\author[0000-0003-1479-3059]{G.~S.~Stringfellow}
\affiliation{University of Colorado Boulder, Boulder CO 80309 USA
}
\email{Guy.Stringfellow@colorado.edu}

\author[0000-0003-4341-6172]{A.~K.~Vivas}
\affiliation{Cerro Tololo Inter-American Observatory/NSF NOIRLab, Casilla 603, La Serena, Chile}
\email{kathy.vivas@noirlab.edu}

\collaboration{30}{(DELVE  Collaboration)}

\begin{abstract}
We present the discovery of a compact Milky Way satellite in the constellation of Gemini. This system was discovered by cross-matching detection\update{s} from two independent search algorithms applied to Blanco/DECam data from \update{the third data release of the DECam Local Volume Exploration survey (DELVE DR3)}, and confirmed with deeper imaging from Gemini/GMOS-N. 
Based on these data, we determine that the system is an ultra-faint (\satmv), compact \CHECK{($r_{1/2} = 8.6^{+1.4}_{-1.2}$\,pc)} system located at a heliocentric distance of $120^{+7}_{-6}$~kpc. These physical properties place the system in the regime of ambiguous, ultra-faint compact Milky Way halo satellites that cannot be confidently classified as dwarf galaxies or star clusters from morphology alone; we therefore name the system \satname. From medium-resolution Keck/DEIMOS spectroscopy, we securely identify four members including two blue horizontal branch stars, confirming the system as a bound satellite moving at a mean radial velocity of $v_{\rm hel} = -82.7\update{^{+3.7}_{-3.9}}$\,\kms. We also use these spectra to place an upper limit of $\rm [Fe/H] \lesssim -2.5$ on the metallicity of \satname's brightest star, supporting the classification of the system as either an ancient star cluster or ultra-faint dwarf galaxy. The discovery of faint, distant systems similar to \satname \  \update{is} expected to become more common with \update{upcoming surveys}.
\end{abstract}


\shortauthors{Overdeck et al.\ (DELVE Collaboration)}
\shorttitle{A New Distant, Compact Milky Way Satellite in Gemini}


\section{Introduction} 

\begin{figure*}
    \centering
    \includegraphics[width=1\textwidth]{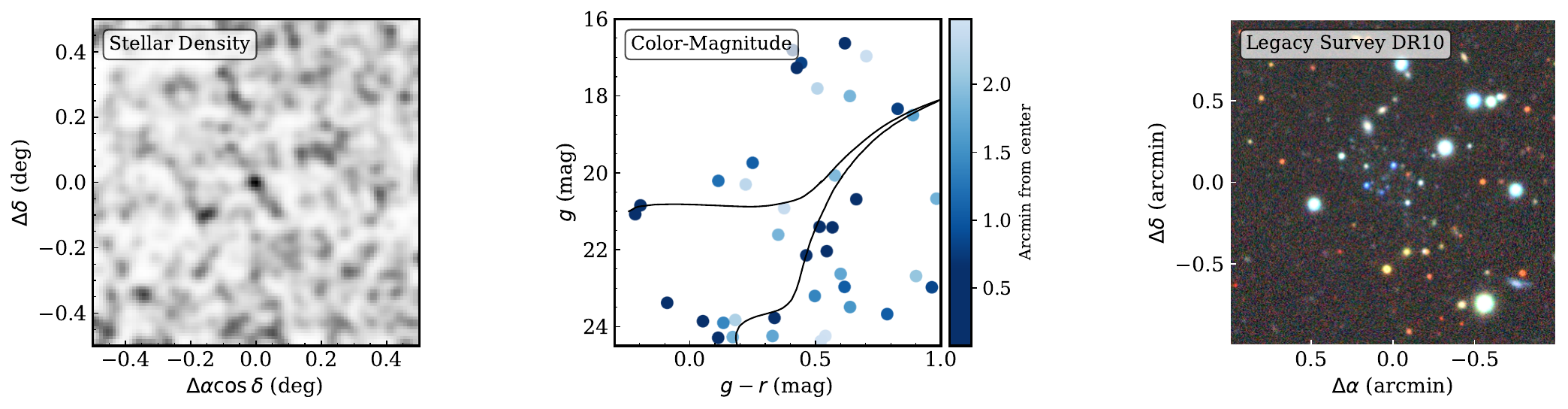}
    
    \caption{\label{fig:simple_plot} 
    Three initial views of \satname{} at the time of its discovery. (Left) Isochrone-filtered stellar density map for a \CHECK{$1^\circ \times 1^\circ$} region centered on \satname, smoothed with a Gaussian kernel with \CHECK{$\sigma = 0.5\arcmin$}.
    (Center) Color--magnitude diagram \update{(CMD)} assembled from the DELVE DR3 photometry with a \CHECK{13.5\,Gyr, $Z = 0.0001$} \texttt{PARSEC} isochrone overplotted. Stars are colored according to their distance from the center of \satname.
    (Right) False-color image of a \CHECK{$2\arcmin \times 2\arcmin$} region centered on \satname from the Legacy Surveys Sky Viewer DR10.}
\end{figure*}

\setcounter{footnote}{0}

Ultra-faint dwarf galaxies (UFDs) represent an extreme in galaxy formation. With absolute magnitudes fainter than $M_V = -7.7$ and stellar masses sometimes extending to less than $10^3\,M_\odot$, they are the least luminous, most dark-matter-dominated, and most chemically pristine stellar systems known \citep{Simon2019,2026arXiv260210202G}. Their extreme properties make them uniquely powerful laboratories: the census of Milky Way satellites constrains the minimum mass required for star formation in dark matter halos \citep[e.g.,][]{Bullock2000, Benson2002, Sommerville2002, Okamoto2008, Nadler2020, Ahvazi2024}, while their internal kinematics probe dark matter physics on spatial scales of tens of parsecs \citep{Bullock2017}. UFDs are also among the most promising targets for indirect searches for dark matter annihilation or decay 
\citep{GeringerSameth2015,McDaniel2024}.
\par The known Milky Way satellite population has grown substantially over the past two decades, 
driven by progressively deeper wide-field surveys, including the 
Sloan Digital Sky Survey \citep[SDSS;][]{Willman2005a, Willman2005b, 
Belokurov2006}, PanSTARRS-1 \citep[PS1;][]{Chambers2016, Laevens2015a, 
Laevens2015b}, the Dark Energy Survey \citep[DES;][]{DES:2016, 
Bechtol2015, Koposov2015}, \update{the DECam Local Volume Exploration Survey} \citep[DELVE;][]{Drlica-Wagner2021,Drlica-Wagner:2022}, the Hyper Suprime-Cam Strategic Survey Program 
\citep[HSC-SSP;][]{Aihara2018, Homma2019, Homma2024} and the Ultraviolet Near Infrared Optical Northern Survey (UNIONS; \citealt{2025AJ....170..324G,2023AJ....166...76S,2024ApJ...961...92S}).
Collectively, these efforts have contributed to a known population of roughly 65 confirmed and candidate Milky Way satellite galaxies \citep{Pace2025}, with many more galaxies predicted \update{to be discovered by ongoing and future surveys} \citep{Nadler2020, Ahvazi2025, 2025OJAp....8E..89T,  Tan2026}.

The boundary between \textit{bona fide} galaxies and star clusters grows increasingly ambiguous at the faintest luminosities (i.e., $M_V \gtrsim -3.5$), where observed systems, including DELVE~1 \citep{Mau2020}, Eridanus~III \citep{Bechtol2015, Koposov2015}, and Ursa Major~III/UNIONS~1 \citep{Smith2024}, challenge both spectroscopic confirmation techniques \citep{Simon2024,2026arXiv260217652C} and the conceptual distinction between galaxies and star clusters \citep{Gutcke:2024ApJ...971..103G, Taylor:2025Natur.645..327T}.
Recent theoretical work suggests that genuine dark-matter-dominated galaxies may persist to extraordinarily low stellar masses \citep{Errani2020, Manwadkar2022, Ahvazi2025, 2026arXiv260409539S}, further motivating \update{efforts to detect and confirm the faintest dwarfs with wide-field surveys}.
\par Here we report the discovery of \satname, an ultra-faint Milky Way satellite located at equatorial coordinates ($\alpha_{2000}, \delta_{2000}) = (118.38\arcdeg,   15.99 \arcdeg$), identified in the  \update{new third DELVE data release  (DELVE DR3; \citealt{Tan:2025c}; Drlica-Wagner et al. in prep.).}
The paper is organized as follows: in Section~\ref{sec:disc} we describe the DELVE DR3 data and our two dwarf detection methods, in Section~\ref{sec:gemini} we describe deeper follow-up imaging with Gemini North \update{t}elescope, and in Section~\ref{sec:keck} we describe results of spectroscopic follow-up with the Keck II Telescope. We discuss the implications of \satname in Section~\ref{sec:discussion}.

\section{Data and Discovery} 
\label{sec:disc}

The DELVE survey  \citep{Drlica-Wagner2021, Drlica-Wagner:2022} is designed to discover and characterize ultra-faint satellites around the Milky Way, the Magellanic Clouds, and isolated Magellanic analogs in the Local Volume. DELVE uses the Dark Energy Camera \citep[DECam;][]{Flaugher2015} on the V\'{i}ctor M.\ Blanco 4-meter Telescope at  NSF's Cerro Tololo Inter-American Observatory (CTIO) in Chile to assemble contiguous imaging of the high-Galactic-latitude southern sky in the $g$, $r$, $i$, and $z$ bands.

The DELVE program has been allocated $>$180 nights of dedicated observations, which are combined with public archival DECam imaging, including data from the Dark Energy Survey \citep[DES;][]{DES:2016}, the DECam Legacy Survey \citep[DECaLS;][]{Dey:2019}, and the DECam eROSITA Survey \citep[DeROSITAS;][]{Zenteno:2025}. DELVE DR3 (\citealt{Tan:2025c}; Drlica-Wagner et al.\ in prep.) provides source catalogs derived from coadded images processed uniformly through the DES Data Management pipeline \citep[DESDM;][]{Morganson:2018}, with image detrending and coaddition following the procedures of DES DR2 \citep{DES:2021}. \update{Through} the coadding process, we achieve a deeper median 10$\sigma$ point-source depth for DELVE DR3 of $g\sim24.2$ and $r\sim23.7$ \citep{Tan2026}, which is a substantial improvement over the depths reported in DR2 of $g\sim23.5$ and $r\sim23.1$ \citep{Drlica-Wagner:2022}.

To identify Milky Way satellite candidates, we search for spatial overdensities consistent with old, metal-poor stellar populations. This task is complicated by the faintness of these systems: their member stars are vastly outnumbered by foreground Milky Way stars and distant background galaxies in the DELVE DR3 catalogs.
We employ two independent algorithms to search for satellite candidates: an isochrone matched-filter approach \citep[\texttt{simple\footnote{https://github.com/DarkEnergySurvey/simple}};][]{Bechtol2015, Drlica-Wagner:2020} and a likelihood-based rasterized search \citep[\texttt{ugali}\footnote{https://github.com/DarkEnergySurvey/ugali};][]{Bechtol2015, Drlica-Wagner:2020}.  The two methods use different techniques to suppress foreground and background contamination and enhance the stellar signal from potential dwarf galaxy candidates. Both show similar sensitivity to real systems, but their false positives are largely uncorrelated. Requiring a candidate to be independently detected by both searches therefore substantially reduces contamination \citep[see Figure~4 of][]{Tan2026}.

The \texttt{simple} algorithm has been used to discover at least 32 ultra-faint Milky Way satellites in DECam data \citep[e.g.,][]{Bechtol2015, Drlica-Wagner:2015,Mau2020, Cerny:2023b, Tan2026b}. It first selects stellar sources in color--magnitude space consistent with an old ($\tau = 12$\,Gyr), metal-poor ($Z = 0.0001$, $[\mathrm{Fe/H}] \sim -2.2$) \texttt{PARSEC v1.2S} isochrone \citep{Bressan:2012, Chen:2014, Tang:2014, Chen:2015} in the $g-r$ bands, and smooths the filtered stellar density field with a $\sigma = 2\arcmin$ Gaussian kernel. It then searches for spatial overdensities among the filtered stellar sample by partitioning the search area into $\texttt{HEALPix}$ pixels with $\texttt{nside} = 32$ and identifying peaks by iteratively raising the density threshold until fewer than 10 disconnected regions remain above threshold. For each overdensity, the algorithm iterates over circular apertures with radii between 0.6$\arcmin$ and 18$\arcmin$ in steps of 0.6$\arcmin$.  \update{It then  calculates the Poisson detection significance, $\texttt{SIG}_{gr}$, relative to the background density for each aperture, and chooses the radius that maximizes $\texttt{SIG}_{gr}$.}

The \texttt{ugali} algorithm takes a complementary, likelihood-based approach. It compares a model incorporating a dwarf galaxy signal (i.e., spatial distribution, color--magnitude distribution in the $g-r$ bands, and initial mass function) against a null model consisting only of a uniform field of foreground stars and misclassified background galaxies estimated empirically from objects in a surrounding annulus ($0.5^\circ < r < 2.0^\circ$), \update{and calculates a test statistic ($\sqrt{\texttt{TS}_{gr}}$)  based on the likelihood ratio between the two models}. The dwarf galaxy model assumes a \citet{Plummer1911} spatial distribution, while the color–magnitude distribution is generated from old, metal-poor \texttt{PARSEC} isochrones populated according to the \citet{Chabrier:2001} initial mass function. The likelihood is constructed by evaluating, for each star within $r < 0.5^\circ$ of a candidate position, the probability of membership based on its spatial position, photometry, and color. Full details of the likelihood construction are given in \citet{Bechtol2015} and Appendix~C of \citet{Drlica-Wagner:2020}.

We run the \texttt{simple} and \texttt{ugali} over the entire DELVE DR3 coadd catalogs, using the same procedure as \citet{Tan2026b}. This search resulted in 439 hotspots that exceeded the detection thresholds of \texttt{simple} ($\texttt{SIG}_{gr} \geq  4.0$) and \texttt{ugali} ($\sqrt{\texttt{TS}_{gr}} \geq 5.0$) and were matched between the two search algorithms to within $0.2^\circ$. We investigated each of the hotspots by generating diagnostic plots similar to Figure~\ref{fig:simple_plot}. \satname emerged as one of the more statistically significant candidates that had not already been identified ($\texttt{SIG}_{gr} = 4.9$, $\sqrt{\texttt{TS}_{gr}} = 5.6$), and was found to be particularly compelling due to the presence of two candidate blue horizontal branch (BHB) member stars. Furthermore, inspection of Legacy Surveys Sky Viewer images coincident with \satname showed a visible overdensity of blue stars consistent with other compact Milky Way satellites (Figure~\ref{fig:simple_plot}).  
While \satname was identified alongside the other systems presented in \citet{Tan2026b}, its confirmation required deeper data that were not yet available when \citet{Tan2026b} was published. Here we present those follow-up observations, which convincingly establish \satname as a new Milky Way satellite.

\section{Gemini/GMOS Deep Imaging: Morphology}
\label{sec:gemini}

\begin{figure}
    \centering    \includegraphics[width=0.90\linewidth]{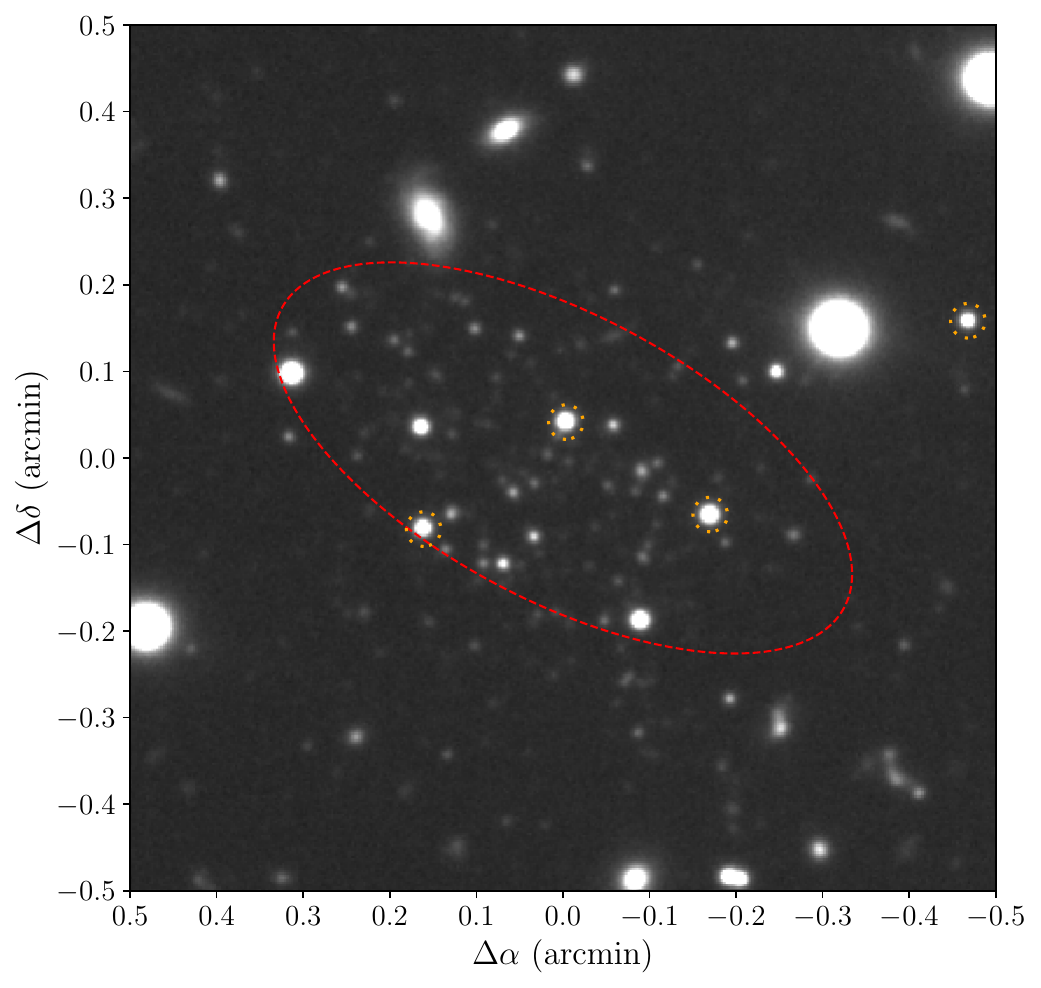}
    \caption{Follow-up $r$-band Gemini/GMOS image of \satname. Relative to the Legacy Survey image (Figure \ref{fig:simple_plot}), the deeper observations reveal an increased number of faint, point-like sources concentrated at the target location, providing additional evidence that \satname is a bona-fide Milky Way satellite.  We also overlay a red ellipse on the image indicating the our best-fit angular half-light radius, $a_h = 0.37\arcmin$, \update{as well as orange circles marking the spectroscopically confirmed member stars of the system.} }
\end{figure}

\begin{figure*}
    \centering
    \includegraphics[width=\textwidth]{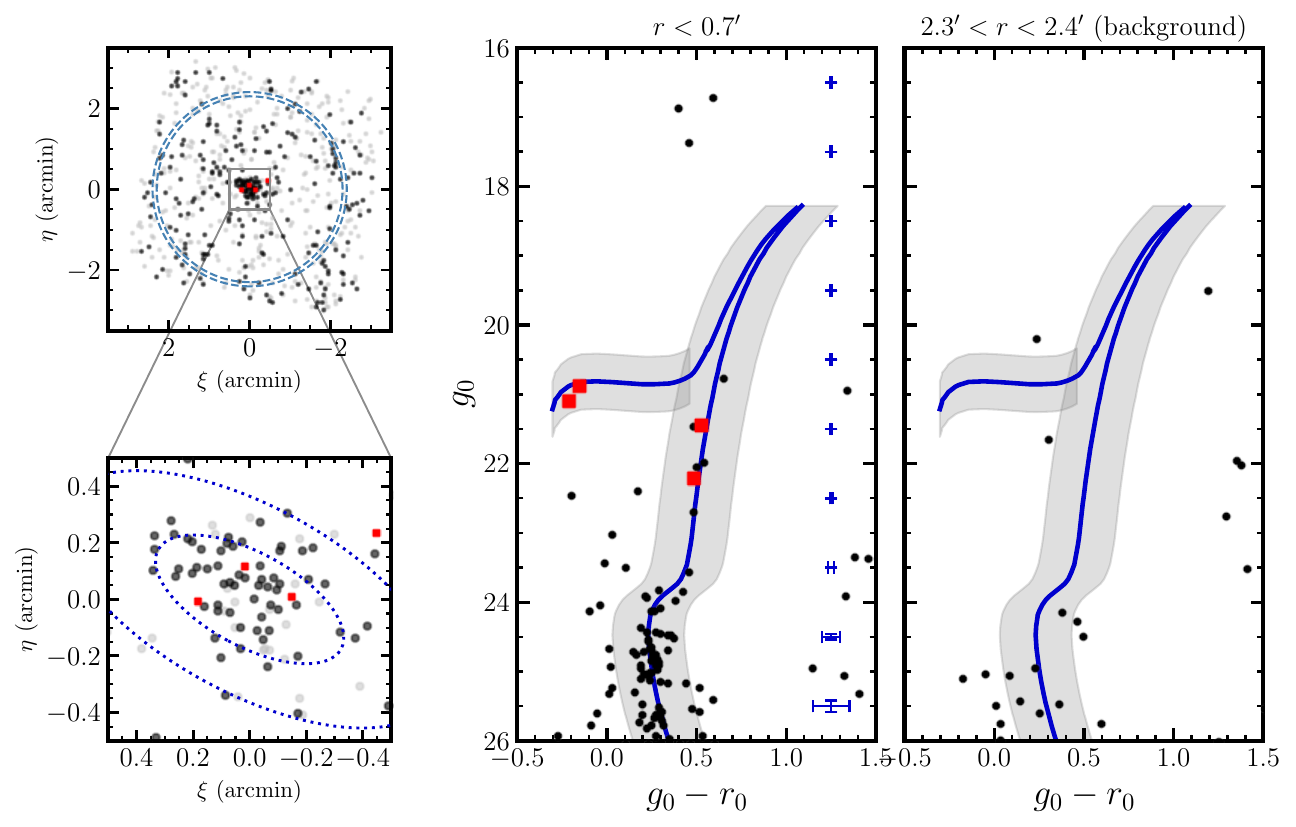}
    \caption{(Left-top) Spatial distribution of all stars in our Gemini/GMOS-N photometry, centered on \satname.  Points in black pass our isochrone filter (see right-hand panels) and were used for the structural fitting, while those in gray correspond to those that fall outside this filter.  The four kinematic members identified with Keck/DEIMOS spectroscopy are red squares. A background control region defined by inner radius $r=2.3\arcmin$ and outer radius $r=2.5\arcmin$ is shown by blue dashed circles; this corresponds to the largest possible annulus fitting within the $5.5\arcmin\times5.5\arcmin$ FOV of GMOS-N. (Left-bottom) Zoom-in of a $0.5\arcmin \times 0.5\arcmin$ region centered on \satname. \update{Dotted ellipses correspond to [1,2]$\times a_h$.} (Center)  \update{CMD} of \satname within a circular selection of $r < 0.7\arcmin$, based on on the GMOS-N photometry. A $\tau = 13.5$\,Gyr, $Z = 0.0001$ \texttt{PARSEC} isochrone is shown in blue, set to our best-fit distance modulus of $(m-M)_0 = 20.39$.  (Right) A similar color--magnitude diagram showing stars within the background annulus depicted in the top-left panel.}
    \label{fig:deepcmd}
\end{figure*}

\subsection{Observations and Photometry}
To confirm the existence of \satname and better constrain its morphological properties, we obtained deep $g,r$-band follow-up imaging with the Gemini Multi-Object Spectrograph at Gemini North (GMOS-N) located on Maunakea. Observations were executed in queue mode in February 2025 and February 2026. Observations in 2025 provided sufficiently deep images to confirm the reality of the system, motivating us to pursue an observing program in 2026 with more stringent data quality requirements (IQ:70\%, CC:50\%, SB:50\%, as defined by the international Gemini Observatory)\footnote{\url{https://www.gemini.edu/observing/telescopes-and-sites/sites\#Constraints}} to constrain the properties of the system by reaching $\sim$2\,mag below the main-sequence turn-off. In total, we obtained 10$\times$150\,s exposures ($g$-band) and 14$\times$150\,s exposures ($r$-band) in clear, dark skies with an average seeing of $0.68\arcsec$ and $0.60\arcsec$, respectively. Small dithers between frames were applied to cover the GMOS chip gaps. All observations used $2\times2$ binning, providing a plate scale of $0.16\arcsec/{\rm pixel}$.

\par We reduced the GMOS images using the Gemini \textsc{dragons} package \citep{2019ASPC..523..321L}\footnote{\url{https://dragons.readthedocs.io/en/stable/}}, which performs overscan and bias subtraction, flat-fielding, image stacking, and cosmic ray rejection based on the stacked images. For these corrections, we used flat field and bias frames  provided by the Gemini Observatory archive that were taken closest in time to our data.

\par We performed point-spread function (PSF) photometry on the coadded GMOS $g$ and $r$ images using the \textsc{daophot iv}/\textsc{allstar} suite \citep{1987PASP...99..191S}. An empirical PSF was derived for each image using $\sim$80--90 non-saturated stars that were spread over the entire field-of-view but not located close to the edges of the CCDs. We adopted a Moffat function ($\beta=2.5$) for the PSF model since it provided the smallest residuals. As in \citet{Martinez-Vazquez:2021a}, we performed initial photometry on the stacked images with a spatially-constant PSF using {\sc allstar}; this provided a preliminary catalog and a star-subtracted image. To recover faint sources that were undetected in the first pass (especially those in the PSF wings of brighter objects), we identified additional stars in the star-subtracted image and appended them to our previous catalog \citep[e.g.,][]{Martinez-Vazquez:2021a, Cantu:2021}. Lastly, after subtracting the PSF stars from the stacked images, we refined the PSF model and allowed a quadratic variation of the PSF across the GMOS FOV in order to perform the final photometry. \update{Throughout the analysis described below, we define stars as sources passing the empirically-determined  selections $-0.5 < \texttt{SHARP} < 0.5$ and $\chi < 3$ based on the \textsc{daophot} outputs.}

\par The GMOS-N instrumental photometry was transformed into the DECam $g,r$ system by matching objects from the GMOS-N catalog to objects in DELVE DR3. Linear transformations from GMOS-N $g,r$ into DECam $g,r$ were derived allowing for a color term to account for differences in the instrumental systems. \update{The rms scatter of the photometric transformations is 0.020 mag in the $g$ band and 0.016 mag in the $r$ band.}

\begin{deluxetable}{lccc}
\tabletypesize{\scriptsize}
\tablewidth{0pt} 
\tablecaption{Measured and derived parameters of \satname. Details of each parameter can be found in their corresponding sections. \label{tab:struct} }
\tablehead{
\colhead{Parameter} & \colhead{Description}& \colhead{Value} & \colhead{Units}   } 
\startdata 
\multicolumn{2}{l}{\textbf{Morphological Fits (Section~\ref{sec:gemini})}}&& \\
$\alpha_{J2000}$& Right Ascension of Centroid & $118.386^{+0.001}_{-0.001}$ & deg \\ 
$\delta_{J2000}$& Declination of Centroid & $15.990^{+0.001}_{-0.001}$ & deg \\ 
$a_h$ &  Angular Semi-Major Axis Length & $0.37^{+0.07}_{-0.06}$ & arcmin  \\
$a_{1/2}$&  Physical Semi-Major Axis Length & $12.8^{+2.7}_{-2.2}$ &  pc  \\ 
$r_h$&  Azimuthally-Averaged Angular Half-Light Radius & $0.25^{+0.04}_{-0.03}$ &  arcmin \\
$r_{1/2}$&  Azimuthally-Averaged Physical Half-Light Radius & $8.6^{+1.4}_{-1.2}$ & pc  \\ 
$\epsilon$&  Ellipticity & $0.6_{-0.1}^{+0.1}$ &  ---  \\
P.A.&  Position Angle of Major Axis (East of North) & $62^{+7}_{-7}$ & deg \\ 
$(m-M)_0$&  Distance Modulus$^a$& $20.39  \update{\pm 0.12}$ & mag\\
$D_\odot$&  Heliocentric Distance$^a$ & $120_{-6}^{+7}$ & kpc \\ 
$\tau$& Age$^b$ & 13.5 & Gyrs  \\
$Z$ &  Isochrone Metallicity$^b$ & 0.0001 & --- \\
$E(B-V)$ & \update{Galactic Extinction}  & 0.030  & ---  \\
\hline
$M_V$&  Absolute (Integrated) $V$-band Magnitude & \satmvnolabel &  mag   \\ 
$L_V$&  $V$-band  Luminosity & $580^{+430}_{-170}$ &  $L_\odot$  \\
$M_*$& Stellar Mass (Assuming $M_*/L_V=2$) & $1150^{+860}_{-330}$ & $M_\odot$  \\
\hline
\multicolumn{2}{l}{\textbf{Properties from Keck/DEIMOS Spectroscopy (Section~\ref{sec:keck})}}&& \\
$N_{\rm spec}$& Number of Spectroscopically Confirmed Members  Stars &  4  &  - \\
$v_{\rm sys}$& Mean Heliocentric Radial Velocity & $-82.7^{+3.7}_{-3.9}$ &  km s$^{-1}$  \\ 
$\sigma_{v}$& Line-of-Sight Velocity Dispersion (95\% credible limit) & $< 8.9$ &  km s$^{-1}$  \\ 
$\rm [Fe/H]_{\rm brightest}$ & Spectroscopic Metallicity of Brightest Star (approximate upper limit) & $\lesssim -2.5$ &  - \\
\enddata
\tablenotetext{a}{Following \citet{Drlica-Wagner:2015}, we added in quadrature a 0.1 mag systematic uncertainty  to the distance modulus measurement to account for the uncertainties in the isochrone modeling.}
\tablenotetext{b}{Age and isochrone metallicity were fixed in our fits to the oldest age and lowest metallicity available in our PARSEC grid \update{when selecting the stellar sample for our distance estimation and morphological fit}.}
\end{deluxetable}

\subsection{Distance}
We determined the distance of \satname by fitting \texttt{PARSEC} isochrones to bootstrap realizations of the observed GMOS color--magnitude diagram. To begin, we selected a sample of stars at $20 < g_0 < 26$ within a circular radial selection of $r<0.7\arcmin$, corresponding to $\sim 3\times$ an initial estimate of the azimuthally-averaged half-light radius of the system. Next, we fit a grid of $\tau = 13.5$ Gyr, $Z=0.0001$ \texttt{PARSEC} isochrones spanning a range \update{of} distance moduli bounded on [20.1, 20.7] in steps of 0.01 to the red giant branch (RGB), main sequence (MS), and blue horizontal branch (BHB) of the resulting color--magnitude diagram. For each distance modulus in this grid, we interpolated the RGB and MS of the isochrone and computed a $\chi^2$ statistic comparing the observed and isochrone-predicted $(g-r)_0$ colors, using per-star color errors with an extra 0.03 mag color spread added in quadrature, i.e., $\sigma_{\rm g-r, tot} = \sqrt{0.03^2 + \sigma_g^2 + \sigma_r^2}$. Stars with color differences $>$0.2 mag from the isochrone were excluded. For the BHB, we instead compared only the $g$-band magnitude difference between the isochrone and the observed BHB stars, with an additional 0.1 mag to account for HB modelling uncertainties, i.e., $\sigma_{\rm g,tot} = \sqrt{0.1^2 + \sigma_g^2}$. The best-fitting distance modulus was taken as the value that yielded the smallest $\chi^2$. 
\par This procedure was repeated across 25,000 bootstrap realizations, and we determined our final distance and uncertainty from the median and 16th/84th percentiles of these realizations. We further added a 0.1 mag systematic uncertainty to account for uncertainties in the isochrone models \citep{Drlica-Wagner:2015}, as well as our choice to fix the age and metallicity during this fitting.
\par We arrived at a distance modulus of \satmod, corresponding to a heliocentric distance of \satheliodist.

\subsection{Structural Parameters}

To measure the structural parameters of \satname, we fit the structural properties using the binned Poisson maximum-likelihood formalism described in Appendix C of \citet{Simon2021} and implemented for GMOS in \citet{Cerny:2023b}. We first defined a stellar sample by selecting all stars in the calibrated photometric catalog consistent with the  \texttt{PARSEC} isochrone set to the distance modulus derived in the previous subsection. Specifically, we  applied a color cut of $\Delta(g-r)_0 < 0.2$ around the MS and RGB of the best-fit isochrone, and a magnitude cut of $\Delta g_0 < 0.4$ about the HB (see gray shaded region in Figure~\ref{fig:deepcmd}). We also applied a bright-end magnitude limit of $g_0 < 20$ to reduce interloper contamination.

\par In addition to the stellar catalog, the likelihood analysis requires a coverage map of the region of sky that is covered by the data. As in \citet{Cerny:2023b}, we first defined a binary map at the full GMOS image resolution by  assigning  all pixels in the stacked GMOS $g$-band image above (below) a fixed threshold of 100 counts a value of one (zero), which isolated the region of the GMOS detector used for imaging. Then, we masked out pixels flagged in the data quality (DQ) array produced by the \textsc{DRAGONS} reduction pipeline, which identifies cosmic rays, saturated pixels, and known bad pixels/columns. Next, we identified and masked regions occupied by bright foreground stars and extended background galaxies from the coverage map. Object identification for this masking was performed with the \texttt{SEP} source extraction package \citep{Barbary2016}, which was used to fit the position, size, and ellipticity of these sources. Finally, we binned this full-resolution map into $25\times25$ pixels and assigned a coverage fraction to each binned pixel.

\par With the two key inputs defined (the stellar sample and the coverage map), we proceeded to fit \satname's morphology. We modeled the stellar density profile of \satname as an elliptical Plummer profile \citep{Plummer1911}. The free parameters in the fit were the total number of observed member stars ($N_*$), the centroid position in pixel coordinates ($X_0$, $Y_0$), the projected semi-major axis  ($a_h$), the ellipticity ($\epsilon = 1-b/a$), the position angle (P.A.) in the image frame, and a uniform background surface density ($\Sigma_b$, in stars per $10^6$ pixels) to account for foreground/background objects that pass the isochrone filter. The predicted model counts in each spatial bin were multiplied by the corresponding value of the binned coverage fraction mask, allowing the likelihood function to accurately account for the non-uniform detector coverage and masked regions when comparing to the observed binned star counts.

We sampled the posterior probability distribution of each free parameter using the affine-invariant MCMC ensemble sampler \texttt{emcee} \citep{ForemanMackey2013}.  We ran the sampler with 100 walkers evolved for 10,000 steps and discarded the first 2,500 as burn-in. To assess the robustness of our structural parameter recovery, we repeated the fit over a range of magnitude limits from $g_0 < 23.0$ to $g_0 < 26.0$ in steps of 0.5\,mag, finding consistent results across all selections that retained a sufficient number of member stars \CHECK{($N \gtrsim 25$)}. The best-fit parameter values converge with smaller error bars at increased depth, with the exception of the background stellar density and the number of observed member stars, which are expected to grow for deeper limiting magnitudes. The shallower cuts ($g_0 \lesssim 24.5$, $N < 25$) yield poorly converged chains. We ultimately chose to report parameter estimates and uncertainties based on the median and 16th/84th percentiles of the marginalized posterior distributions derived with a magnitude limit of $g_0 < 25$; at fainter limits, we found that star/galaxy separation became more challenging and the stellar completeness begins to drop.
\par The azimuthally-averaged half-light radius was derived from the semi-major axis and ellipticity via $R_h = a_h\sqrt{1-\epsilon}$, and all parameters were converted from pixel to angular units and subsequently to physical units at the best-estimated distance of \satname (\satheliodist). The resulting structural parameters are listed in Table~\ref{tab:struct}.

\subsection{Absolute Magnitude}
We determined the stellar mass and absolute magnitude \update{from the GMOS-N data} following a standard simulation-based approach \update{\citep[e.g.,][]{Martin2008, Cerny:2023b}}. Adopting the best-fit isochrone age, metallicity, and distance modulus determined in the preceding subsections, we simulated mock stellar populations  across a logarithmic grid of stellar masses $[10 \rm \ M_\odot, 10^4 \ M_\odot]$ assuming a Chabrier initial mass function. For each trial stellar mass, we generated 1000 mock populations and computed the number of \update{stars brighter than}  a magnitude limit of $g_0 = 25$ (the magnitude limit used for our structural fit) and compared it to the number of isochrone-filtered stars \update{that were observed} when constrained via a structural fit above the same \update{magnitude} limit ($N_*\approx49$). From the model that generated a mean number of stars that is closest to the observed data, we determined the richness \update{as an estimate of the total number of stars down to the hydrogen-burning limit \citep{Bechtol2015, Drlica-Wagner:2020}}. Finally, the absolute magnitude was determined following the stellar population sampling procedure described by \citet{Martin2008} as implemented in \texttt{ugali}. We inspected various diagnostics such as the simulated CMDs and the cumulative luminosity function of stars, confirming the strong resemblance between our mock populations and the observed data.\footnote{\update{For example, we found that the mass-matched simulated populations typically hosted 0-3 BHB stars and often lacked RGB stars brighter than the HB level.}}
\par From this procedure, we determined an absolute magnitude of \satmv{} for \satname{}, corresponding to a $V$-band luminosity of \satlv. \update{The uncertainty on these estimates primarily arises from the variation in the number of bright stars across the mock populations owing to the stochastic sampling of the initial mass function; we neglect the uncertainty on \satname{}'s distance.} Adopting a stellar mass-to-light ratio of $\Upsilon = 2 \ M_*/L_V$ as expected for an old
stellar population \citep{Simon2019}, this translates to \satmstar. 

\section{Keck Spectroscopy} 
\label{sec:keck}

\begin{figure*}
    \centering
    \includegraphics[width=0.95\textwidth]{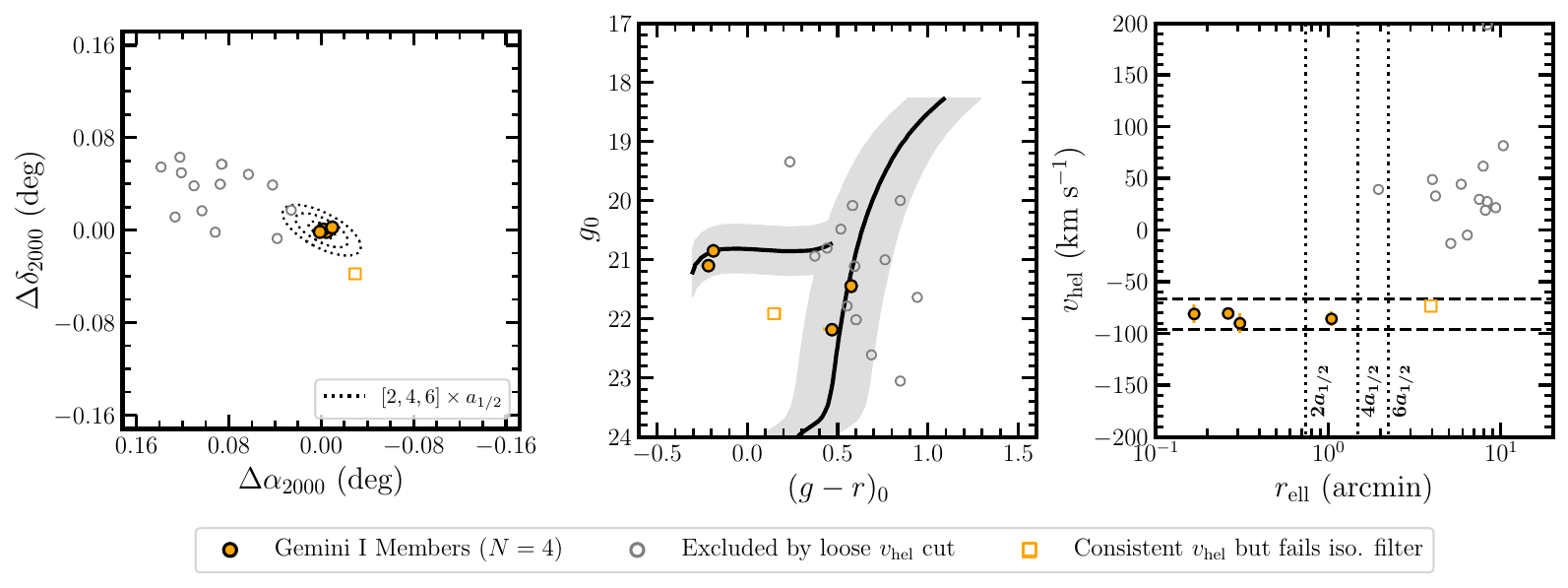}
    \caption{(Left) Spatial distribution of stars with measured velocities from Keck/DEIMOS. Likely member stars are shown as orange circles, one velocity-consistent non-member is shown as an orange square, and all other non-member stars are shown as gray circles. Contours of $[2,4,6] \times a_{1/2}$ are shown as black dotted ellipses. (Center) Color--magnitude diagram of the same targets; we use DELVE DR3 photometry here because some spectroscopic targets fall outside the GMOS-N FOV. In black, we overplot a $\tau =13.5$ Gyr, $Z=0.0001$ \texttt{PARSEC} isochrone; the grey shading corresponds to our membership selection of $0.2$~mag about this model RGB and 0.4 mag about the model BHB. (Right) Velocity ($v_{\rm hel}$) vs.\ projected elliptical radius ($r_{\rm ell}$). The dotted vertical lines correspond to $[2,4,6] \times a_{1/2}$, matching the left panel. Velocity errorbars are shown but are often smaller than the marker symbol size. \label{fig:spectra}}
\end{figure*}

\begin{deluxetable*}{ccccccccc}
\label{table:keck_members}
\tablecaption{Properties of spectroscopically confirmed member stars of \satname, in addition to one velocity-consistent non-member. Stars are ordered by Keck/DEIMOS spectrum S/N (highest to lowest).}
\tablehead{
    \update{Star Name} & RA & DEC & $g_0$ & $r_0$ & S/N & $v_{\rm hel}$ & Type & Member \\
    & (deg) & (deg) & (mag) & (mag) & & (\kms) & &
}
\startdata
     Gaia DR3 666966265968082304 & $118.38205$ & $15.98891$ & $21.44 \pm 0.02$ & $20.87 \pm 0.02$ & $9.5$ & $-80.3 \pm 3.5$ & RGB & True \\
    Gaia DR3 666966364750007552 & $118.38494$ & $15.99069$ & $20.85 \pm 0.02$ & $21.04 \pm 0.02$ & $6.0$ & $-81.1 \pm 9.0$ & BHB & True \\
    \update{J075330.45+155933.51} & $118.37689$ & $15.99264$ & $22.17 \pm 0.03$ & $21.71 \pm 0.03$ & $4.6$ & $-85.7 \pm 7.3$ & RGB & True \\
    \update{J075333.07+155919.08} & $118.38779$ & $15.98863$ & $21.10 \pm 0.02$ & $21.32 \pm 0.02$ & $4.4$ & $-89.8 \pm 9.5$ & BHB & True \\
    \hline
    \update{J075325.54+155709.05} & $118.35644$ & $15.95251$ & $21.91 \pm 0.11$ & $21.76 \pm 0.03$ & $4.1$ & $-73.4 \pm 7.5$ & \ldots & False \\
\enddata
\end{deluxetable*}

\subsection{Observations and Data Reduction}
    We obtained medium-resolution, multi-object spectroscopy of \satname with the DEep Imaging Multi-Object Spectrograph (DEIMOS; \citealt{2003SPIE.4841.1657F}) mounted on the 10-m Keck II telescope at the W.M. Keck Observatory on Maunakea, Hawai`i.\footnote{\update{The raw spectra were retrieved from the Keck Observatory Archive (\url{https://koa.ipac.caltech.edu}; program Y295; PIs: Cerny, Geha) and will become publicly available after a period of 18 months.}} These observations used the 1200G grating with central wavelength 7700\,\AA, providing a spectral resolution $R \approx 6000$ across the wavelength range $\sim$6500--9000\,\AA. We observed a single slitmask for the system comprised of $\sim$34 slits of width 0.7\arcsec and minimum length 4.5\arcsec. This mask included the 5 most likely member stars in the system based on a preliminary \texttt{ugali} fit; two remaining brighter member candidates could not be targeted due to slit collisions.  The total exposure time was 2.2~hours split across four exposures, all taken under cloudy conditions with poor ($\gtrsim 1\arcsec$) seeing. 

\par We reduced the DEIMOS data, measured stellar velocities, and determined equivalent widths (EWs) of the Calcium Triplet (CaT) lines following the procedures described extensively in \citet{2026ApJ...999..140G}. Briefly, the raw 2D spectra were reduced with \texttt{PypeIt} \citep{2020JOSS....5.2308P}, which performs standard reduction operations including flat fielding, sky subtraction, and spectral extraction. Velocities were determined from the extracted 1D spectra using the \texttt{dmost} package\footnote{\url{https://github.com/marlageha/dmost}}, which forward-models each spectrum with both a synthetic spectral template from the PHOENIX library \citep{2013A&A...553A...6H}, as well as a telluric template from \texttt{TelFit} \citep{2014AJ....148...53G}. \update{To determine CaT EWs, each line was fit with a Gaussian-plus-Lorentzian profile (for $S/N > 15$ spectra) or a Gaussian profile ($S/N < 15$, or cases where the Gaussian-plus-Lorentzian fit did not converge).}

\subsection{Stellar Membership and Internal Kinematics}
The DEIMOS observations described above yielded a spectroscopic catalog containing 29 stars with measured velocities. We present a spatial distribution map, color--magnitude diagram, and velocity vs.\ projected-radius plot in the left, center, and right panels of Figure~\ref{fig:spectra}, respectively. The kinematic signal of \satname is clearly identifiable in the latter panel as an overdensity of $N=4$ stars at $v_{\rm hel} \approx -85\pm5$~\kms. One additional star at $r_{\rm ell} \approx 6a_{1/2}$ has a similar velocity but falls outside our isochrone selection window in both DELVE DR3 and in our GMOS-N photometry; it is therefore more likely to be a non-member. We summarize the properties of these five stars in Table~\ref{table:keck_members}.

\par The consistency of these stars' velocities adds convincing evidence that \satname is a bound, co-moving association of stars. However, with just four member stars and relatively poor per-star velocity precision (3.5--10\,\kms), no meaningful constraints on \satname's internal kinematics -- and therefore its dark matter content and classification -- are possible. Nonetheless, we estimated \satname's mean velocity ($v_{\rm sys}$) by modeling its velocity distribution through a two-parameter Bayesian fit constraining both $v_{\rm sys}$ and the system's intrinsic velocity dispersion $\sigma_v$; the latter serves as a nuisance parameter in this case \citep[e.g.,][]{2006AJ....131.2114W}. Adopting flat priors of [$-500$\,\kms, $500$\,\kms] for $v_{\rm  sys}$ and [0\,\kms, 10\,\kms] for $\sigma_v$, we performed dynamic nested sampling with \texttt{dynesty} \citep{2020MNRAS.493.3132S}. From the resulting marginalized posteriors, we find a mean velocity of $v_{\rm sys} = -82.7^{+3.7}_{-3.9}$\,\kms, with a 95\% credible upper limit of $\sigma_v < 8.9$\,\kms for the velocity dispersion that is largely set by the adopted prior.  We summarize these estimates in Table~\ref{tab:struct}.

\par Owing to the lack of \textit{Gaia} proper motions for member stars in \satname---either among our spectroscopic sample or among the photometric member catalogs from \texttt{ugali}---this velocity measurement provides limited immediate insight into \satname's orbit. However, we note that two of these spectroscopic members have position and brightness information in  \textit{Gaia} DR3 (see Table~\ref{table:keck_members}), and we expect that future \textit{Gaia} data releases may enable study of \satname's orbital history based on its 6D kinematics. 

\par Lastly, we attempted to derive an Equivalent-Width (EW)-based Calcium Triplet (CaT) metallicity for \satname{}'s brightest giant ($S/N \approx 9.5$) using \texttt{dmost}. We found that noise around the $\rm 8498 \ \AA$  and $\rm 8662 \ \AA$ lines precluded a standard estimate based on all three lines. However, we were able to recover a reasonable fit to the strongest line, for which we found an EW of $\sim$0.8$\ \rm \AA$. Taking $3\times$ this EW as an upper bound on the summed CaT EW across all three lines suggests a total EW of $< 2.4 \rm \ \AA$ for the star. Translated to a metallicity using the calibration from \citet{2026MNRAS.546ag019N} following the procedure from \citet{2026arXiv260217652C}, this suggests a metallicity upper limit for \satname{}'s brightest star of $\rm [Fe/H]_{\rm brightest} \lesssim -2.5$.

\begin{figure*}[t!]
    \centering
    \includegraphics[width=\textwidth]{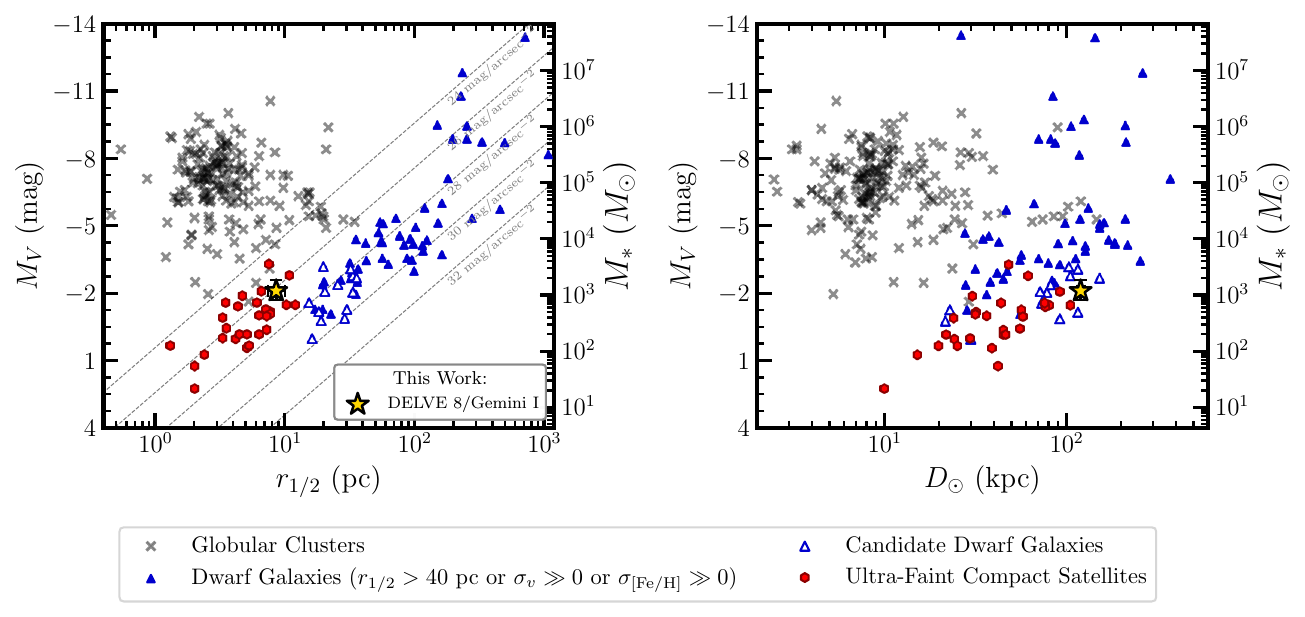}
    \caption{\label{fig:mvrhalf}
    Comparing the properties of \satname to the population of Milky Way globular clusters (black $\times$s), candidate and confirmed dwarf galaxies (unfilled and filled blue triangles, respectively) and ultra-faint compact satellites (red hexagons).   (Left) In the $M_V$--$r_{1/2}$ plane, \satname's properties are consistent with the population of ultra-faint compact satellites ($r_{1/2} < 15$~pc; $\mu > 24 \rm \ mag \ arcsec^{-2}$, $M_V > -3.5$), which have proven challenging to classify.
    (Right) In the $M_V$--$D_{\odot}$ plane, \satname is more distant than nearly all Milky Way globular cluster population and is the most distant ultra-faint compact satellite yet discovered.}
\end{figure*}

\section{Discussion and Final Remarks} 
\label{sec:discussion}
\update{We have presented the discovery of \satname:  an ultra-faint (\satmv), compact ($r_{1/2} = 8.3$~pc) stellar system in the outer Milky Way halo ($D_\odot = 122~{\rm kpc}$)}. Posed in terms of \textit{apparent} magnitude, \satname ($m_V = 18.3$) is \CHECK{the fourth faintest satellite yet discovered by the DELVE survey} (after DELVE 5, Phoenix~III, DELVE~7, at $m_V = [18.4,19.1,19.3]$, respectively), reflecting the increasingly sensitive searches now possible with the deeper DELVE DR3 dataset \citep{Tan2026b,2026ApJ..1000...87T}.\par Morphologically, \satname's properties place it in a particularly ambiguous region of the size-luminosity parameter space: as seen in Figure \ref{fig:mvrhalf}, it is smaller than any confirmed Milky Way satellite galaxy, yet fainter than nearly all known classical globular clusters. DELVE 8/Gemini I is thus the latest  member of the Ultra-Faint Compact Satellites (UFCSs) -- a designation for the diverse group of satellites at $M_V>-3.5; r_{1/2} < 15$~pc that likely includes both the faintest star clusters and the least-massive galaxies \citep{2026arXiv260217652C}. For example, the \satname{} system exhibits morphological properties similar to Eridanus III ($M_V = -2.1$; $r_{1/2} = 6.5~{\rm pc}$, after taking the azimuthal average of the result from \citealt{2018ApJ...852...68C}). Eridanus III is considered a galaxy candidate: its chemical properties are suggestive of a galactic nature but its dark matter content remains ambiguous \citep{2018ApJ...852...68C,2023ApJ...958..167F,Simon2024,2026arXiv260217652C}. To reflect this current classification ambiguity, we have named the system \satname following the convention that star clusters are typically named after the survey they are discovered within while satellite galaxies are named after their host constellation and order of discovery therein.  

\par Classifications in this regime are best distinguished through spectroscopic measurement of a velocity or metallicity dispersion \citep{2012AJ....144...76W}. Unfortunately, neither \update{of these measurements is} currently possible from our DEIMOS spectroscopy due to the low signal-to-noise and small number of member stars. Nonetheless, an important clue to \satname's classification comes from the system's ancient age and our spectroscopic metallicity upper limit. In particular, isochrone fits to the  blue horizontal branch and main-sequence turnoff disfavor ages $<10$~Gyr, largely excluding the possibility that \satname is a younger/intermediate-age globular cluster resembling those that have been found among the faintest satellites of  the Galactic halo (see e.g., \citealt{1993ASPC...48...38Z,1998AJ....115..648R,2007A&A...466..181C,2013MNRAS.433.1966O,2014AJ....148...19P,2016ApJ...822...32W}  and the discussion in \citealt{2026arXiv260217652C}). This is independently supported by the low spectroscopic metallicity upper limit of $\rm [Fe/H]_{brightest} < -2.5$ (derived in Section~\ref{sec:keck}), which suggests the system is more consistent with either ancient globular clusters at the metal-poor end of the metallicity distribution, or, alternatively, an ultra-faint dwarf galaxy on the empirical luminosity-metallicity relation \citep[e.g.,][]{2013ApJ...779..102K,2026arXiv260210202G}. These properties strongly motivate deeper spectroscopy of the system, which holds the promise of detecting or ruling out a metallicity spread among member giants.

\par In pursuit of additional clues about \satname{}'s nature and origins, we examined the Local Volume Database \citep{Pace2025} and the \texttt{galstreams} package \citep{Mateu2023MNRAS.520.5225M} to search for associations with other MW structures. Of note, \satname is $\sim10^{\circ}$ away from Koposov~2 \citep{Koposov2007ApJ...669..337K}, another ambiguous ultra-faint compact satellite and the only previously known ultra-faint MW satellite in the Gemini constellation. There are no known stellar streams that overlap with \satname, but the Sagittarius stream is relatively nearby. 
In terms of Sagittarius stream coordinates, \satname has ($\Lambda$, $B$) = (191.84$^{\circ}$, 14.37$^{\circ}$) using the coordinate system from \citet{Majewski2003ApJ...599.1082M}.
As the Sagittarius stream can have an extent of $\sim10^{\circ}$, and the near ($D_\odot\approx 20 ~{\rm kpc}$) and far ($D_\odot\approx$75--160~{\rm kpc}) arms spatially overlap in this region, there is a potential association. However, the  N-body models of \citet{2021MNRAS.501.2279V} suggest that the distant arm has a heliocentric velocity of $\sim60~{\rm km~s^{-1}}$ at $\Lambda$ = 191.84$^{\circ}$. 
Our velocity measurement of $-82.7~{\rm km~s^{-1}}$ thus rules out any association with the Sagittarius stream, and the origins of \satname{} remain unknown.

Looking ahead, the Vera C.\ Rubin Observatory's Legacy Survey of Space and Time will soon deliver photometry several magnitudes deeper than DELVE DR3 over a comparable sky area \citep{2019ApJ...873..111I}, dramatically increasing the census of faint stellar systems in the Local Group and beyond \citep{2025OJAp....8E..89T}. The discovery and characterization of systems like \satname, which reside at the boundary of our current capabilities, is therefore timely: developing the tools and frameworks to study these objects will be essential for interpreting the forthcoming flood of discoveries from LSST. Even in the eventuality that individual systems remain ambiguous -- including in cases where spectroscopy is available -- population-level analyses still promise to unveil a new understanding about the threshold of galaxy formation.

\section*{Acknowledgments}
After our identification and initial follow-up of this candidate, we became aware that \satname had been independently identified by McQuinn, Mao, Buckley, Shih, Dolphin, Cohen, Tollerud, Hai, Leishman, and Brown as a candidate Local Group galaxy (\citetalias{McQuinn:2025hst..prop18066M}).

\par \update{KO was supported from the NSF grant AST-2006340}. WC gratefully acknowledges support from a Gruber Science Fellowship at Yale University.  This material is based upon work supported by the National Science Foundation Graduate Research Fellowship Program under Grant No.\ DGE2139841. CYT was supported by the U.S.\ National Science Foundation (NSF) through grants AST-2108168 and AST-2307126. DJS acknowledges support from NSF grant AST-2205863. DC acknowledges support from NSF grant AST-2508747. BMP acknowledges support from NSF grant AST-2508745.  
\par The DELVE project is partially supported by the NASA Fermi Guest Investigator Program Cycle 9 No.\ 91201 and by Fermilab LDRD project L2019-011. This material is based upon work supported by the National Science Foundation under Grant No.\  AST-2108168, AST-2108169, AST-2307126, and AST-2407526. This research award is partially funded by a generous gift of Charles Simonyi to the NSF Division of Astronomical Sciences.  The award is made in recognition of significant contributions to Rubin Observatory’s Legacy Survey of Space and Time.  Any opinions, findings, and conclusions or recommendations expressed in this material are those of the author(s) and do not necessarily reflect the views of the National Science Foundation.
DJS acknowledges support from NSF grant AST-2508746.

Funding for the DES Projects has been provided by the U.S. Department of Energy, the U.S. National Science Foundation, the Ministry of Science and Education of Spain, 
the Science and Technology Facilities Council of the United Kingdom, the Higher Education Funding Council for England, the National Center for Supercomputing 
Applications at the University of Illinois at Urbana-Champaign, the Kavli Institute of Cosmological Physics at the University of Chicago, 
the Center for Cosmology and Astro-Particle Physics at the Ohio State University,
the Mitchell Institute for Fundamental Physics and Astronomy at Texas A\&M University, Financiadora de Estudos e Projetos, 
Funda{\c c}{\~a}o Carlos Chagas Filho de Amparo {\`a} Pesquisa do Estado do Rio de Janeiro, Conselho Nacional de Desenvolvimento Cient{\'i}fico e Tecnol{\'o}gico and 
the Minist{\'e}rio da Ci{\^e}ncia, Tecnologia e Inova{\c c}{\~a}o, the Deutsche Forschungsgemeinschaft and the Collaborating Institutions in the Dark Energy Survey. 

The Collaborating Institutions are Argonne National Laboratory, the University of California at Santa Cruz, the University of Cambridge, Centro de Investigaciones Energ{\'e}ticas, 
Medioambientales y Tecnol{\'o}gicas-Madrid, the University of Chicago, University College London, the DES-Brazil Consortium, the University of Edinburgh, 
the Eidgen{\"o}ssische Technische Hochschule (ETH) Z{\"u}rich, 
Fermi National Accelerator Laboratory, the University of Illinois at Urbana-Champaign, the Institut de Ci{\`e}ncies de l'Espai (IEEC/CSIC), 
the Institut de F{\'i}sica d'Altes Energies, Lawrence Berkeley National Laboratory, the Ludwig-Maximilians Universit{\"a}t M{\"u}nchen and the associated Excellence Cluster Universe, 
the University of Michigan, NSF NOIRLab, the University of Nottingham, The Ohio State University, the University of Pennsylvania, the University of Portsmouth, 
SLAC National Accelerator Laboratory, Stanford University, the University of Sussex, Texas A\&M University, and the OzDES Membership Consortium.

Based in part on observations at NSF Cerro Tololo Inter-American Observatory at NSF NOIRLab (NOIRLab Prop. ID 2012B-0001; PI: J. Frieman, NOIRLab
Prop. ID 2019A-0305; PI: Alex Drlica-Wagner, which is managed by the Association of Universities for Research in Astronomy (AURA) under a cooperative agreement with the National Science Foundation.

The DES data management system is supported by the National Science Foundation under Grant Numbers AST-1138766 and AST-1536171.
The DES participants from Spanish institutions are partially supported by MICINN under grants PID2021-123012, PID2021-128989 PID2022-141079, SEV-2016-0588, CEX2020-001058-M and CEX2020-001007-S, some of which include ERDF funds from the European Union. IFAE is partially funded by the CERCA program of the Generalitat de Catalunya.

We  acknowledge support from the Brazilian Instituto Nacional de Ci\^encia
e Tecnologia (INCT) do e-Universo (CNPq grant 465376/2014-2).

This document was prepared by the DES Collaboration using the resources of the Fermi National Accelerator Laboratory (Fermilab), a U.S. Department of Energy, Office of Science, Office of High Energy Physics HEP User Facility. Fermilab is managed by Fermi Forward Discovery Group, LLC, acting under Contract No.\ 89243024CSC000002.

Based in part on observations obtained at the international Gemini Observatory (Prop. ID GN-2026A-Q-138; PI: Martínez-Vázquez), a program of NSF NOIRLab, which is managed by the Association of Universities for Research in Astronomy (AURA) under a cooperative agreement with the U.S. National Science Foundation on behalf of the Gemini Observatory partnership: the U.S. National Science Foundation (United States), National Research Council (Canada), Agencia Nacional de Investigaci\'{o}n y Desarrollo (Chile), Ministerio de Ciencia, Tecnolog\'{i}a e Innovaci\'{o}n (Argentina), Minist\'{e}rio da Ci\^{e}ncia, Tecnologia, Inova\c{c}\~{o}es e Comunica\c{c}\~{o}es (Brazil), and Korea Astronomy and Space Science Institute (Republic of Korea). Gemini data were acquired through the Gemini Observatory Archive \citep{2017ASPC..512...53H} at NSF NOIRLab and processed using DRAGONS (Data Reduction for Astronomy from Gemini Observatory North and South, \citealt{2019ASPC..523..321L}, Zenodo DOI: 10.5281/zenodo.17412281). 

\par This work was enabled by observations made from the Gemini North telescope and from the Keck II telescope. The scientific community is honored to have the opportunity to conduct astronomical research on Maunakea in Hawai‘i. We recognize and acknowledge the very significant cultural role and reverence of Maunakea to the Kanaka Maoli (Native Hawaiians) community.

\par Some of the data presented herein were obtained at Keck Observatory, which is a private 501(c)3 non-profit organization operated as a scientific partnership among the California Institute of Technology, the University of California, and the National Aeronautics and Space Administration. The Observatory was made possible by the generous financial support of the W. M. Keck Foundation.

\par This research has made use of the Keck Observatory Archive (KOA), which is operated by the W. M. Keck Observatory and the NASA Exoplanet Science Institute (NExScI), under contract with the National Aeronautics and Space Administration.

\par This work has made use of data from the European Space Agency (ESA)
mission {\it Gaia} (\url{https://www.cosmos.esa.int/gaia}), processed by the {\it Gaia} Data Processing and Analysis Consortium (DPAC,
\url{https://www.cosmos.esa.int/web/gaia/dpac/consortium}). Funding
for the DPAC has been provided by national institutions, in particular
the institutions participating in the {\it Gaia} Multilateral Agreement.


\bibliography{main}{}
\bibliographystyle{aasjournalv7}



\end{document}